\documentstyle[11pt,newpasp,twoside, epsf]{article}
\def\edcomment#1{\iffalse\marginpar{\raggedright\sl#1\/}\else\relax\fi}
\reversemarginpar
\begin{document}
\title{Starburst-AGN Connection from High Redshift to the Present Day}
\author{Yoshiaki Taniguchi}
\affil{Astronomical Institute, Graduate School of Science,
              Tohoku University, Japan}

\begin{abstract}
We give a review on possible starburst-AGN (active galactic nuclei)
connections from high redshift to the present day.
First, we give a historical review on some basic ideas related to
the starburst-AGN connection published in the literature.
Second, we focus our attention to the so-called Magorrian
relation which is the close relationship between the nuclear black hole
mass  and the bulge mass, established in nearby galaxies.
If the Magorrian relation is universal, we obtain an important
implication that any supermassive black holes were made through
successive merging processes of starburst remnants (i.e., neutron
stars and stellar-sized black holes), providing a channel of
the starburst-AGN connection.
Third, we briefly discuss a possible new scenario for the formation of
quasar nuclei at very high redshift
based on an idea that the successive mergers
of starburst remnants formed in subgalactic gaseous clouds.
\end{abstract}


\section{General Introduction}

\subsection{Introduction}

Nuclear (or circumnuclear) gas is often ionized at some level
in most nearby galaxies (in particular, disk galaxies).
It is known that there are two fundamental types of nuclear activity;
(1) the nuclear starburst activity and (2) the nonthermal nuclear activity.
Nuclear gas in the former class of galaxies is photoionized by massive OB stars
while than in the latter class of galaxies is photoionized by nonthermal
ionizing continuum emission from the central engine of active galactic
nuclei (AGN) (e.g., Rees 1984).
According to the recent extensive spectroscopic study of nuclear regions of 486
nearby galaxies promoted by Ho, Filippenko, \& Sargent (1997),
the both types of galactic nuclei share approximately 40\%, respectively,
if we include objects with low emission-line luminosity or low activity.
However, typical luminous starburst nuclei share approximately several
percent of nearby galaxies (e.g., Balzano 1983). Also, Seyfert nuclei,
typical AGNs in  nearby universe, are found in approximately 10 percent of
nearby galaxies (Ho et al. 1997).
Therefore, roughly speaking, $\approx$ 10\% of galactic nuclei experience
the nuclear starburst activity, $\approx$ 10\% of galactic nuclei experience
the nonthermal activity, and the remaining $\approx$ 80\% of galactic nuclei
show little evidence for significantly high level of such activities
(i.e., nearly normal galactic nuclei, hereafter NGNs).

Here a question arises as; ^^ ^^ Why do some galactic nuclei experience the nuclear
starburst ? Why do some galactic nuclei have an AGN ? Why do the majority of
nearby galaxies show little evidence for such activities 
in their nuclear regions ?"
These fundamental questions can be replaced by a fascinating question;
^^ ^^ Are there any evolutionary connections among the three types of galactic nuclei ?"
If this is the case, we have further important questions;
^^ ^^ How are they connected ? What are important physical processes
in such connections ?"
Indeed, many astronomers, including the authors, have been enslaved
by the so-called starburst-AGN connection. We will introduce main ideas
proposed up to now in next section.

\subsection{Proposed Starburst-AGN Connections in the Literature}

Although many ideas on the starburst-AGN connections have been
proposed up to now, they may be broadly classified as follows.

\begin{description}

\item{[1]} {\it From starburst to AGN through the formation of a supermassive
black hole}: This idea suggests that a supermassive black hole (SMBH), which is
believed as the key ingredient of the central engine of AGNs, is made
through successive mergers among starburst remnants (e.g., Weedman 1983;
Norman \& Scoville 1988; Taniguchi, Ikeuchi, \& Shioya 1999; Ebisuzaki
et al. 2001; see also Taniguchi et al. 2002b; Mouri \& Taniguchi 2002b).

\item{[2]} {\it From starburst to AGN due to the starburst-driven gas fueling
onto a supermassive black hole}: This idea suggests that the gas fueling can be
supplied either from gaseous envelope of supergiant stars near the supermassive
black hole (Scoville \& Norman 1988), from supernova ejecta (Taniguchi 1992),
or from the gas associated with the nucleus of a merging partner
(Taniguchi 1999; see also Taniguchi \& Wada 1996).

\item{[3]} {\it From starburst to AGN-like phenomena}: This idea is completely
different from the above two ideas because the photoionization of nuclear gas
is attributed to some descendents of massive stars; e.g., hot Wolf-Rayet
stars (Warmers: Terlevich \& Melnick 1985), supernovae in dense gas media
(Terlevich et al. 1992), shock heating by superwinds (Heckman 1980; Taniguchi 1987;
Taniguchi et al. 1999), or hot planetary nebula nuclei (Taniguchi, Shioya, \&
Murayama 2000a; see also Shioya et al. 2002). Note that the shock heating is
though to work in some LINERs (=Low Ionization Nuclear Emission-line Regions)
and ULIRGs (= Ultraluminous Infrared Galaxies, or ULIGs).
See also Rieke et al. (1988)
for a possible evolutionary path of a nuclear starburst.

\item{[4]} {\it From ULIRGs to quasars}: This idea suggests an evolutionary
link between ULIRGs and quasars; i.e., ULIRGs are precursors of quasars
in the local universe
(Sanders et al. 1988). In this model, mergers between two or more gas-rich galaxies
are crucially important to initiate very luminous nuclear starbursts
in the central region of the merger remnant (see also Taniguchi \& Shioya 1998;
Taniguchi 1999).

\item{[5]} {\it From  ULIRGs and/or LIRGs through S2s to S1s}:
In this idea, type 2 Seyferts (S2s) are considered as a possible
missing link between ULIRGs and/or LIRGs (=luminous infrared galaxies)
and type 1 Seyferts (S1s) (Heckman et al. 1989; Mouri \& Taniguchi 1992, 2002a).
The reason for this is that S2s tend to have circumnuclear starburst regions
more often than S1s
and their starburst ages appears older than those of typical nuclear starbursts
(e.g., Cid Fernandes et al. 2001; Storchi-Bergman et al. 2001;
Mouri \& Taniguchi 2002a and references therein).

\end{description}

Although all the above ideas may not always work in actual galaxies,
it seems better to keep in mind the following points.
(a) Massive stars formed in a nuclear starburst evolve through
hot phases (i.e., Wolf-Rayet stars, planetary nebula nuclei, and so on)
to supernova explosions inevitably. Therefore, we have to take account of
all the evolutionary phases when we discuss the evolution of starburst nuclei.
(b) Compact remnants (i.e., stellar-sized black holes and neutron stars)
are also inevitably remained in the nuclear starburst region. Therefore,
we have to think about the dynamical evolution of such remnants
under a realistic gravitational potential together with dynamical
interactions with existing stars in the concerned region.
Careful consideration on these two points makes it possible to
discuss the starburst-AGN connection.

\subsection{Toward a Simple Unified Model for Triggering AGNs}

As outlined briefly, there are some
possible evolutionary connections among 
AGNs, starburst galactic nuclei (SGNs), and
NGNs.  Prior going to discussion on the
starburst-AGN connection, let us consider why some galactic nuclei
are AGNs in which the central SMBH plays an
important role. If only galaxies with an AGN could have a SMBH in their
nucleus, it would be easily understood why some galaxies have AGNs.
However, recent high-resolution optical spectroscopy of a sample
of nearby, normal galaxies have shown that most nucleated galaxies
have a SMBH in their center (e.g., Richstone et al. 1998; Magorrian et
al. 1998). Furthermore, the relationship between the SMBH mass and
the bulge (or spheroid) mass is basically similar for AGNs and
NGNs (Gebhardt et al. 2000; Ferrarese et al. 2001;
Wandel 2002). Therefore, the presence of a SMBH in the nucleus is not
a crucial discriminator between AGNs and NGNs.

The frequency of occurrence of luminous AGNs (i.e., Seyfert nuclei)
in nearby galaxies (i.e., $\approx$ 10\%) implies that a typical lifetime
of such nuclear activity is $\sim 10^9$ yr. Therefore, it is
suggested strongly that some NGNs could be triggered to evolve to
AGNs and then die after a duration of  $\sim 10^9$ yr. The dead nuclei
should be regarded again as NGNs. From this point of view, it seems reasonable
to imagine that SGNs may provide a missing link between AGNs and NGNs.
We think that this is indeed the starburst-AGN connection which we
want to understand.
In order to explore the whole evolutionary links among AGNs, SGNs,
and NGNs, we have to take account of both the dynamical structure
of host galaxies of all types of nuclei and the environmental effect
(see Figure 1). 

\begin{figure}
\plotone{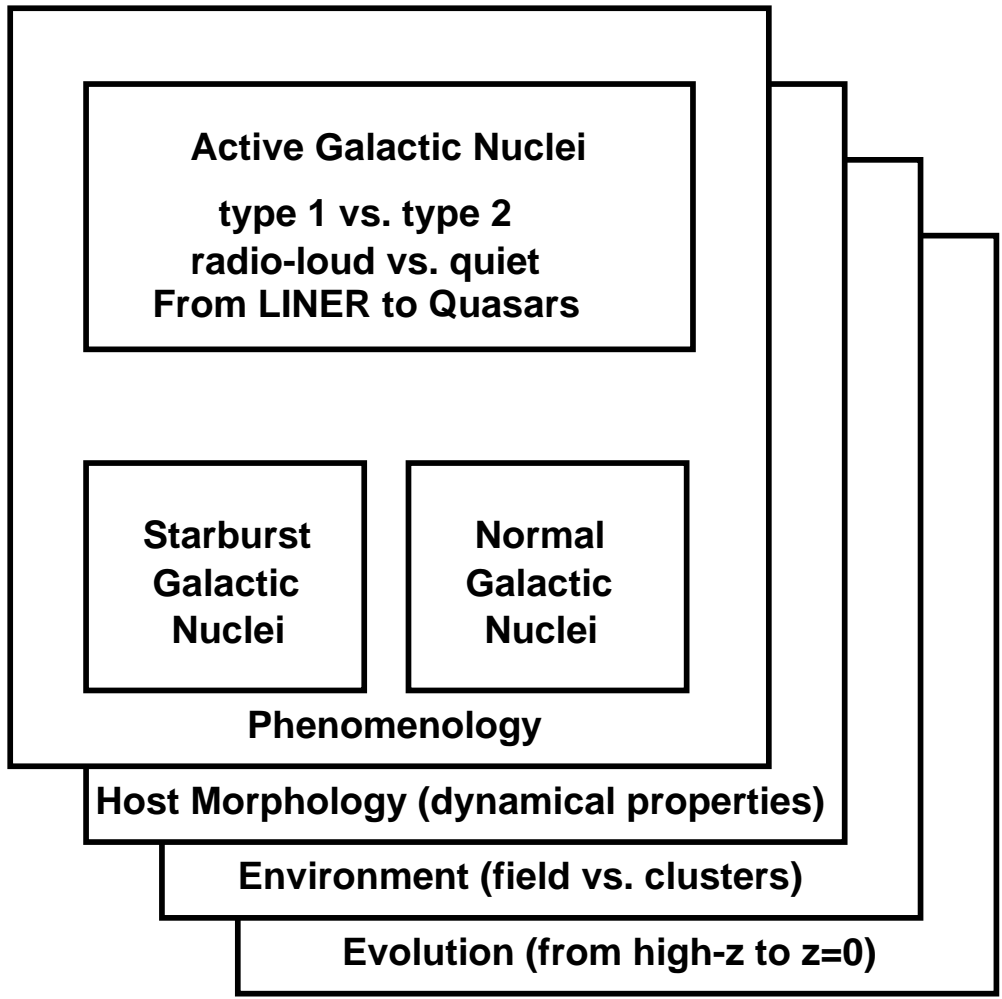}
\caption{A grand unified model toward the understanding
of starburst-AGN connections from high redshift to the present day.}
\end{figure}


Recent systematic studies for large samples of Seyfert nuclei have shown that;
1) Seyfert nuclei do not prefer barred galaxies as their hosts
(e.g.,  Mulchaey \& Regan 1997; Hunt et al. 1999),
and 2) only $\simeq$ 10\% of Seyfert galaxies have companion galaxies
(e.g., De Robertis, Hayhoe, \& Yee 1998a; De Robertis et al. 1998b).
Therefore, it is suggested that both the dynamical effect by non-axisymmetric
structures and the interaction with a companion galaxy give no simple
triggering mechanism for AGNs.
On the other hand, it is quite likely that  any galaxies have been experiencing
minor mergers during their lives (e.g., Ostriker \& Tremaine 1975; Tremaine 1981;
see also Zaritsky et al. 1997). 
Accordingly, Taniguchi (1999) suggested that
the minor-merger-driven fueling appears consistent with
almost all important observational properties of Seyfert galaxies
(see also Taniguchi \& Wada 1996).
Nucleated (i.e., either a SMBH or a dense nuclear star cluster) galaxies and
satellites seem necessary to ensure that the gas in the host disk is surely
fueled into the very inner region (e.g., $\ll$ 1 pc).
Taking account that local quasars may be formed by major mergers between/among
galaxies (e.g., Sanders et al. 1988; Taniguchi \& Shioya 1998;
Taniguchi et al. 1999a), we may have a simple unified formation mechanism
of AGN in the local universe; i.e., all AGNs in the local universe are
triggered by minor or major mergers between/among (nucleated) galaxies,
including satellite galaxies.

\section{Lesson from the Magorrian Relation}

As briefly introduced in Section 1.3, the
recent high-resolution optical spectroscopy of a sample
of nearby, normal galaxies have shown that most nucleated galaxies
have a SMBH in their center (e.g., Richstone et al. 1998; Magorrian et
al. 1998); i.e., the Magorrian relation, hereafter MR.
The most important point of MR in the context of starburst-AGN
connection is that  the ratio between the SMBH mass ($M_\bullet$) and
the bulge (or spheroid) mass ($M_\circ$), $\approx 0.001$ -- 0.002,
 is basically similar for AGNs and
NGNs\footnote{The Magorrian relation is now considered as the relationship
between $M_\bullet$ and the central velocity dispersion of the spheroidal
system (e.g., Tremaine et al. 2002 and references therein). However,
for simplicity, we use the original MR between $M_\bullet$ and $M_\circ$.}
(Gebhardt et al. 2000; Ferrarese et al. 2001;
Wandel 2002). Therefore, we note again that the presence of a SMBH in the nucleus
is not a discriminator between AGNs and NGNs.
Namely, this implies that AGN phenomena are not associated with
the formation process of a SMBH itself.

Recently, Merrifield et al. (2000) investigated a relationship
between the $M_\bullet$/$M_\circ$ ratio and the age of the spheroidal
system for a sample of nearby galaxies and found
that galactic bulges with younger stellar populations
tend to have smaller $M_\bullet$/$M_\circ$ ratios. This suggests that
MR is slightly affected by the recent past starburst.
Since the age spread in their sample galaxies is over several Gyr,
one may estimate the growth timescale of a SMBH, $\tau_\bullet \sim$ 1 Gyr.
However, the above tendency appears weak and thus the $M_\bullet$/$M_\circ$
ratio can be regarded as constant for galaxies with various bulge ages.

In summary, what we have learned from MR can be summarized as follows.

\begin{description}

\item{[1]} Since MR means that the $M_\bullet$/$M_\circ$ ratio is
almost constant for many nearby galaxies, it is suggested that the
SMBH formation is linked physically to the spheroidal formation.
A natural implication seems that a SMBH comes from the coalescence of
nuclear star clusters which formed at the major epoch of spheroidal
formation; i.e., a SMBH comes from mergers of compact remnants
of massive stars born in the spheroidal formation.

\item{[2]} The same MR is found for both AGNs and NGNs. This suggests
that AGN phenomena are not directly linked to the SMBH formation.
Another implication is that the triggering process is much more
important to turn on the nuclear activity.

\item{[3]} The $M_\bullet$/$M_\circ$
ratio appears to be constant for galaxies with various bulge ages,
suggesting that the MR is approximately universal for many galaxies
with the spheroidal component.

\item{[4]} The longer growth timescale of a SMBH ( $\tau_\bullet \sim$ 1 Gyr)
suggests that SMBHs in AGNs may not grow up through the gas accretion
process which helps the mass growth to some extent. This also results
in the same implication as that of Item [1].

\end{description}

As mentioned before, MR provides us some important implications
for the understanding of starburst-AGN connection. Here, adopting
a working hypothesis that MR is universal for galaxies from
high redshift to the present day, we further discuss what MR means.

Suppose that new gas fueling to the spheroidal component of a galaxy
occurred at epoch $t_1$ with the fueled gas mass, $\Delta M_\circ$.
At this epoch, a SMBH with mass of $M_\bullet(t_1)$ was present
in the galaxy center. We observe this system at epoch $t_2$.
The spheroidal mass and the SMBH mass are $M_\circ(t_2)$ and
$M_\bullet(t_2)$, respectively.  At this epoch, we assume that
the SMBH grows up in mass, $\Delta M_\bullet$.  Therefore, we have
the following two relations.

\begin{equation}
M_\bullet(t_2) = M_\bullet(t_1) + \Delta M_\bullet,
\end{equation}
and,

\begin{equation}
M_\circ(t_2) = M_\circ(t_1) + \Delta M_\circ.
\end{equation}
If MR is universal, we find

\begin{equation}
{{M_\bullet(t_2)} \over {M_\circ(t_2)}}
\approx {{M_\bullet(t_1)} \over {M_\circ(t_1)}}.
\end{equation}
In order to achieve the condition given in equation (3),
there are two alternative cases.

\begin{description}

\item{[1]} $\Delta M_\bullet \ll M_\bullet(t_1)$, and
$\Delta M_\circ \ll M_\circ(t_1)$.

\item{[2]} $\Delta M_\bullet \gg M_\bullet(t_1)$, and
$\Delta M_\circ \gg M_\circ(t_1)$.

\end{description}

In the first case, the mass increases in
both the SMBH and the spheroidal component is negligibly small
compared to their original masses.
This case may be applicable to new gas supply onto the nucleus of
a typical disk galaxy in which a SMBH already exists;
i.e., the starburst-AGN connection for
SGNs and AGNs like Seyfert nuclei.

On the other hand, in the second case, the mass increases in
both the SMBH and the spheroidal component are significantly
larger than their original masses.
The conditions [2] require the following relation,

\begin{equation}
{{M_\bullet(t_2)} \over {M_\circ(t_2)}}
\approx {{\Delta M_\bullet} \over {\Delta M_\circ}}
\equiv f_{\rm BH}.
\end{equation}
This case may be applicable both to the ULIRG-quasar connection
at low and intermediate redshift and to the forming galaxy-quasar
connection at high redshift.
It is noted that the universal MR means $f_{\rm BH} \simeq 0.001$ -- 0.002
from high redshift to the present day.
Nice examples for this case are ULIRGs which are believed to be
made from mergers between or among gas-rich galaxies. The nearest
ULIRG, Arp 220, has a number of nuclear super star clusters (SSCs).
Such SSCs will fall into the nuclear region via
the dynamical friction within $\sim 10^{7 -9}$ yr (Shaya et al. 1994;
Taniguchi et al. 1999a), probably making a SMBH with $M_\bullet \sim
10^9 M_\odot$ (see also Norman \& Scoville 1988; Ebisuzaki et al. 2001;
Mouri \& Taniguchi 2002b). Indeed, based on their elaborate numerical
simulations, Bekki \& Couch (2001) found  $f_{\rm BH} \simeq 0.003$
for the aftermath of ultraluminous starburst occurred in a ULIRG.

The formation of SMBHs has been a long standing problem
in modern astrophysics (Rees 1978, 1984). It is still
uncertain how such SMBHs could be born in the nucleus of high-redshift
quasars up to $z \sim 6$. However, recent high-resolution X-ray imaging
studies have discovered
possible candidates of intermediate-mass black holes (IMBHs) with masses
of $M_\bullet \sim 10^{2-4} M_\odot$ in circumnuclear regions
of many (disk) galaxies (e.g., Colbert \& Mushotzky 1999; 
Matsumoto \& Tsuru 1999; Makishima et al. 2000; Strickland et al.
2001; Zezas \& Fabbiano 2002).
It is known that a large number of
massive stars are formed in a circumnuclear giant H {\sc ii} region.
Therefore, Taniguchi et al. (2000b) proposed
 that a continual merger of compact remnants left from
these massive stars is responsible for the formation of such an
IMBH within a timescale of $\sim 10^9$ yr.
A necessary condition is that several hundreds of massive
stars are formed in a compact region with a radius of a few parsecs.
Then Ebisuzaki et al. (2001) proposed that the runaway merging is
an important dynamical process in such a star cluster; its timescale
may be as short as $\sim 10^7$ yr (see also
Mouri \& Taniguchi 2002b). They also proposed  that
circumnuclear star clusters themselves could merge into one during
the course of dynamical evolution in the galaxy potential
within a timescale of $\sim 10^9$ yr.

This idea can be applied to the formation of SMBHs in the hearts of
quasars at high redshift. The number density of quasars peaks at
$z \simeq 2$. Since the growth timescale of a SMBH with $M_\bullet \sim
10^9 M_\odot$ may be $\sim 10^9$ yr, it is required that episodic
massive star formation could occur $\sim 10^9$ yr before $z \simeq 2$;
i.e., $z_{\rm SF} \sim 15$. Even in this dark age, subgalactic gas clumps
with mass of $\sim 10^{7-8} M_\odot$ could form in the context of
cold dark matter scenarios (e.g., Gnedin \& Ostriker 1997).
If numerous subgalactic gas clumps were formed  in a localized region
with a dimension of $\sim$ 10 kpc, piling up compact remnants could make
it possible to form a SMBH with mass of $\sim 10^9 M_\odot$.
According to the universal MR, the spheroidal mass of a quasar host
is $\sim 10^{12} M_\odot$, being comparable to massive galaxies
in the present day.
Although recent optical deep imaging surveys have
shown that bulge formers at high redshift (e.g., Lyman break galaxies
at $z \sim$ 2 -- 4) are very small systems and thus it is unlikely that
they are massive galaxies. On the other hand, recent submillimeter
deep surveys have revealed that massive galaxies are really present
at high redshift although they are basically hidden in the optical
because of heavy extinction by a lot of dust grains (e.g., Frayer et al. 1998, 1999;
Genzel et al. 2002).
Indeed, a high redshift quasar BR 1202$-$0725 at
$z=4.7$ is associated with a massive gaseous system with mass of
$\sim 10^{11 - 12} M_\odot$ (Ohta et al. 1996; Omont et al. 1996).
The first ULIRG beyond $z=2$, IRAS F10214+4724, is also a very massive
system (Downes et al. 1992).
Therefore, since it is likely that precursors of high-redshift quasars
are dust enshrouded massive objects, we can apply the universal
MR for such high-redshift quasars.

\section{Formation of Quasar Nuclei at High Redshift}

Now let us consider
the formation of quasars at high redshift.
In the nearby universe, quasars are associated with nuclei of
(giant) galaxies (e.g., Bahcall et al. 1997).
Therefore, another interesting issue is to investigate a
causal relationship between quasars and galaxies in the
early universe (e.g., Ikeuchi 1981;
Ostriker \& Cowie 1981; Turner 1991; Silk \& Rees 1998; Madau \& Rees 2001).
Here we give an outline of new scenario  for the formation of quasar nuclei (i.e.,
supermassive black holes with mass of $\sim 10^8 M_\odot$)
at high redshift ($z \approx$ 2 -- 5) proposed by Ikeuchi \& Taniguchi (2002).

\begin{description}

\item{Step I}: The formation of subgalactic gas clumps occurs at
$z \approx 15$
as predicted by cold dark matter models.
We assume that the total mass of the clump is $\sim 10^8 M_\odot$ and
the gas mass is $\sim 2\times 10^7 M_\odot$.
Approximately one thousand clumps located within a radius of
10 kpc will be used to build up a galaxy
with a mass of $\sim 10^{11} M_\odot$.

\item{Step II}: The gravitational instability in  clumps could lead to
the formation of massive stars in them. Given the star formation
efficiency of 50\%, $\sim 10^6$ stars with a mass of 10 $M_\odot$
are formed in each clump. These stars ionize about one third
of the gas in the clump.

\item{Step III}: All these massive stars evolve within a timescale of
$\sim 10^7$ yr and then explode as supernovae (SNe).
These SNe overlap each other and then blow out as a superbubble.
Superbubbles arisen from one thousand clumps also overlap
and then evolve into one huge superbubble. This superbubble
can expand at a radius of $\sim$ 500 kpc within a duration of
$\sim 5 \times 10^8$ yr. Since this radius is larger than
the mean separation among galaxies, the IGM is completely
ionized by these superbubbles; i.e., the reionization of
the universe. They also contribute to the metal enrichment up to a level of
$Z \sim 0.01 Z_\odot$.

\item{Step IV}: Compact remnants left from massive stars after the supernova
explosions can merge into one within a duration of $\sim 10^9$ yr.
This leads to the formation of a seed SMBH with a mass of
$\sim 2 \times 10^6 M_\odot$.

\item{Step V}: Approximately fifty seed SMBHs located within a
radius of 500 pc of the galaxy merge into one within
a duration of $\sim 10^9$ yr. Thus a SMBH with a mass
of $\sim 10^8 M_\odot$ is made a few $10^9$ yr after the
initial starbursts in the subgalactic clumps.
This means that quasar nuclei (i.e, SMBHs with $M_{\rm BH}
\sim 10^8 M_\odot$) can be made at $z \approx$ 2 -- 5.

\end{description}

\vspace{0.3cm}

We would like to thank my colleagues, in particular, Satoru Ikeuchi, Hideaki Mouri,
Yasuhiro Shioya, Takashi Murayama, Tohru Nagao, Youichi Ohyama,
and Neil Trentham for useful discussion.


\end{document}